\pdfoutput=1
\documentclass[10pt,a4paper]{article}
\usepackage[T2A]{fontenc}
\usepackage[utf8]{inputenc}
\usepackage[english]{babel}
\usepackage{amssymb,graphicx}
\usepackage{amsmath,amsfonts}
\usepackage{amsthm}
\usepackage{framed}
\usepackage{xcolor}
\usepackage{fullpage}
\usepackage[makeroom]{cancel}
\usepackage[export]{adjustbox} 
\usepackage{microtype}
\usepackage{hyperref}

 \renewcommand{\Im}{\mathop{\mathrm{Im}}\nolimits}
 \renewcommand{\Re}{\mathop{\mathrm{Re}}\nolimits}
 
 \newcommand{\Tr}{\mathop{\mathrm{Tr}}\nolimits}

\theoremstyle{remark}

\numberwithin{equation}{section}

\title{\textbf{Wall crossing, string networks\\
    and quantum toroidal
    algebras}} \author{Yegor~Zenkevich\thanks{yegor.zenkevich@gmail.com}
  \footnote{On leave from ITMP MSU.}\\
  {\small \textit{School of Mathematics, University of Edinburgh,
      UK}}} \date{}
\begin{document}
\maketitle
\vspace{-46ex}
\noindent
\textit{To T.}

\vspace{40ex}

\begin{abstract}
  We investigate BPS states in $4d$ $\mathcal{N}=4$ supersymmetric
  Yang-Mills theory and the corresponding $(p,q)$~string networks in
  Type IIB string theory. We propose a new interpretation of the
  algebra of line operators in this theory as a tensor product of
  vector representations of a quantum toroidal algebra, which
  determines protected spin characters of all framed BPS states. We
  identify the $SL(2,\mathbb{Z})$-noninvariant choice of the coproduct
  in the quantum toroidal algebra with the choice of supersymmetry
  subalgebra preserved by the BPS states and interpret wall crossing
  operators as Drinfeld twists of the coproduct. Kontsevich-Soibelman
  spectrum generator is then identified with Khoroshkin-Tolstoy
  universal $R$-matrix.
\end{abstract}

\section{Introduction}
\label{sec:introduction}
BPS sector of the space of states in a supersymmetric theory is
protected against quantum corrections and can be analyzed even in the
strong coupling regime. The hallmark of this analysis is that the set
of BPS states exhibits intricate discontinuities at certain
codimension one subspaces (walls) in the space of parameters of the
theory. The discontinuities are captured by various wall-crossing
formulas~\cite{KS}. A particularly interesting picture arises in
four-dimensional $\mathcal{N}=4$ supersymmetric gauge theory, where at
low energies at a generic point of the vacuum moduli space the gauge
group $G$ is spontaneously broken to its maximal torus $T$, and the
BPS particles are $W$-bosons, monopoles and dyons charged electrically
and magnetically under $T$.

For $G=U(N)$ one gets an intuitive picture of these states by viewing
the gauge theory as the worldvolume theory on a stack of $N$ parallel
D3 branes in Type IIB string theory. The vacuum moduli space of the
gauge theory
is identified with the configuration space of the parallel D3 branes
in six dimensions transverse to their worldvolume. If all branes are
separated the gauge group is broken to $U(1)^N$ with each $U(1)$ gauge
theory living on a separate D3 brane. Type IIB string theory besides
the fundamental strings (denoted F1 or $(1,0)$) also supports D1 (or
$(0,1)$) branes and an infinite number of $p$F1--$q$D1 bound states
which are known as $(p,q)$ strings. BPS particles in $\mathcal{N}=4$
gauge theory correspond to trivalent networks of $(p,q)$ strings
stretching between D3 branes. The charges as well as tensions of the
strings need to be balanced at every junction. The $p$ (resp.~$q$)
charge of a string ending on a given D3 brane is equal to the electric
(resp.\ magnetic) charge of the BPS particle under the $U(1)$ gauge
group living on the D3 brane. We restrict ourselves to the case when
D3 branes are separated only in two out of six transverse directions
which we denote by $\mathbb{R}^2_{xy}$, and all the $(p,q)$ string
networks are planar. We consider a (twisted) compactification of the
string theory with ``time'' direction running over a circle $S^1$, so
that the partition function is equal to the weighted trace over the
space of BPS states (known as the index, or protected spin
character). The overall ten-dimensional string theory background is
summarized in Table~\ref{tab:1} and an example of a string network is
shown in Fig.~\ref{fig:1}. We provide more details about $(p,q)$
string networks in sec.~\ref{sec:string-networks-type}.
\begin{table}[h]
  \centering
\begin{equation}
    \begin{array}{l|cc|c|c|ccc}
      &\multicolumn{2}{c|}{\text{picture}}&&\text{``time''}&&&\\
      \text{Brane}& \mathbb{R}_x&\mathbb{R}_y&\mathbb{R}_\tau&S^1&\mathbb{C}_{\mathfrak{q}}&\mathbb{C}_{\mathfrak{t}^{-1}}&\mathbb{C}_{\mathfrak{t}/\mathfrak{q}}\\
      \hline
      \mathrm{F1}  &-&&\tau_{*}&-&&&\\
      \mathrm{D1}  &&-&\tau_{*}'&-&&&\\
      \hline
      \mathrm{D3} &x_i&y_i&-&-&--&&
\end{array}\notag
\end{equation}
\caption{Type IIB string theory setup describing $4d$ $\mathcal{N}=4$
  gauge theory living on $\mathbb{C}_{\mathfrak{q}}\times \mathbb{R}_{\tau} \times
  S^1$ and BPS particles in it. The labels on $\mathbb{C}_{\mathfrak{q}}$,
  $\mathbb{C}_{\mathfrak{t}^{-1}}$ and $\mathbb{C}_{\mathfrak{t}/\mathfrak{q}}$ indicate that these
  directions are multiplied (twisted) by $\mathfrak{q}$, $\mathfrak{t}^{-1}$ and $\mathfrak{t}/\mathfrak{q}$
  respectively when going around the $S^1$ ``time'' circle.
  In the ``classical'' limit $q \to 1$ $i$-th D3 brane
  sits at a fixed position $(x_i, y_i)$ in the $\mathbb{R}^2_{x,y}$
  plane, while F1 and D1 (or $(1,0)$ and $(0,1)$ respectively) strings
  may have different slopes (see
  Eq.~\eqref{eq:3} and comments around it); here for concreteness we assume $\Re \tau =
  \alpha = 0$ so that F1 is horizontal and D1 is vertical. Each string
  network sits at a given point $\tau_{*} \in \mathbb{R}_{\tau}$, the
  corresponding operators are ordered by the value of their $\tau_{*}$'s.}
  \label{tab:1}
\end{table}

Type IIB string theory enjoys $S$-duality --- an $SL(2,\mathbb{Z})$
symmetry which leaves the D3 branes invariant, transforms the charge
vectors of $(p,q)$ strings as a two-dimensional vector and $\tau$
using the fractional linear transformations. This symmetry descends to
the gauge theory where it becomes the famous Montonen--Olive
electro-magnetic duality~\cite{MO}, and underlies the
physical approach to the geometric Langland correspondence~\cite{KW}. It will also be crucial in our analysis of line
operators and wall-crossing below.

Our approach rests on the observation that the setup in
Table~\ref{tab:1} is precisely of the form that was related to the
representation theory of quantum toroidal algebras in~\cite{Z1, Z2,
  Z3}. Indeed, the D3 branes are known to correspond to so-called
vector representations $\mathcal{V}_{\mathfrak{q}}$ of the quantum
toroidal algebra
$U_{\mathfrak{q},\mathfrak{t}}(\widehat{\widehat{\mathfrak{gl}}}_1)$
(see Appendix~\ref{sec:prop-repr-quant} for the details about the
algebra and representations). A stack of $N$ D3 branes then
corresponds to a tensor product\footnote{As we will see in
  sec.~\ref{sec:mystery-coproduct} one needs to be careful when
  defining tensor products of representations of a quantum toroidal
  algebra since there is an infinite family of different coproducts. }
$(\mathcal{V}_{\mathfrak{q}}^{*})^{\otimes N}$. Let us describe this
correspondence in some detail. Vector representation
$\mathcal{V}_{\mathfrak{q}}$, as described in
sec.~\ref{sec:vect-repr}, is a representation of
$U_{\mathfrak{q},\mathfrak{t}}(\widehat{\widehat{\mathfrak{gl}}}_1)$
by $\mathfrak{q}$-difference operators of the form $\mathbf{x}^m
\mathbf{y}^n $ with $n,m \in \mathbb{Z}$, and
\begin{equation}
  \label{eq:8}
  \mathbf{y} \mathbf{x} = \mathfrak{q}\, \mathbf{x}
  \mathbf{y}.
\end{equation}
These operators can be understood as elements of the algebra of
functions on a quantum torus with noncommutativity parameter
$\mathfrak{q}$. In the brane picture $\mathfrak{q}$-difference
operator $\mathbf{x}^m \mathbf{y}^{-n}$ corresponds to an endpoint of
a $(n,m)$ string on a D3 brane, for example
\begin{equation}
  \label{eq:6}
  \includegraphics[valign=c]{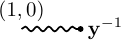}
\end{equation}
where the picture is drawn in the $\mathbb{R}^2_{xy}$ plane, the dot
denotes the D3 brane and the string is a wavy line. The composition of
operators $\mathbf{x}^l \mathbf{y}^{-k} \mathbf{x}^m \mathbf{y}^{-n}$
naturally corresponds to an $(n,m)$ string ending on a D3 brane at
$\tau = \tau_{*}$ and a $(k,l)$ string ending on the same D3 brane at
some $\tau = \tau_{*}' > \tau_{*}$:
\begin{equation}
  \label{eq:7}
    \includegraphics[valign=c]{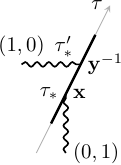}
\end{equation}
where we have attempted a three-dimensional picture representing
$\mathbb{R}^2_{xy}\times \mathbb{R}_{\tau}$. A system of $N$ D3 branes
corresponds to a direct sum of $N$ algebras generated by
$\mathbf{x}_i$, $\mathbf{y}_i$, $i=1,\ldots, N$, so that
\begin{equation}
  \label{eq:16}
  \mathbf{y}_i \mathbf{x}_j = \mathfrak{q}^{\delta_{i,j}}\, \mathbf{x}_j
  \mathbf{y}_i, \qquad [\mathbf{x}_i, \mathbf{x}_j] = [\mathbf{y}_i,
  \mathbf{y}_j] = 0, \qquad i,j =1, \ldots,N.
\end{equation}

In fact, $\mathbf{x}$ and $\mathbf{y}$ play the role of the
coordinates of a D3 brane in the $\mathbb{R}_{xy}^2$ plane. More
precisely, in the limit $\mathfrak{q} \to 1$ the operators
$\mathbf{x}$ and $\mathbf{y}$ commute and should be identified with
complexified exponentiated coordinates\footnote{One needs to be more
  careful when considering framed BPS states. In that case the limit
  will involve the phase parameter $\zeta$ of the line operator.}:
\begin{align}
  \label{eq:9}
  \mathbf{x} &\stackrel{\mathfrak{q}\to 1}{\to} e^{R x +
    i\phi_e},\\
  \mathbf{y} &\stackrel{\mathfrak{q}\to 1}{\to} e^{R y+i\phi_m},
\end{align}
where $\phi_e$ and $\phi_m$ are Wilson and 't Hooft loops of the
$U(1)$ gauge field living on the D3 brane around $S^1$ of radius
$R$. In this ``semiclassical'' limit the intuitive
pictures~\eqref{eq:6},~\eqref{eq:7} with point-like D3 branes in the
$\mathbb{R}^2_{xy}$ plane are actually valid.

For $\mathfrak{q}\neq 1$ the ``coordinates'' $\mathbf{x}$,
$\mathbf{y}$ are non-commutative and hence the position of a D3 brane
cannot be fixed in both $x$ and $y$ directions. One can choose a
\emph{polarization,} i.e.\ a representation of the quantum torus
algebra as difference operators in a single variable, e.g.\
$\mathbf{x}$ (see Appendix~\ref{sec:vect-repr}). Then it is natural to
consider wavefunctions of D3 branes with definite values of
$\mathbf{x}$. It is more appropriate to draw such D3 branes as
vertical dashed lines in the $\mathbb{R}^2_{xy}$ plane rather than
points since the $\mathbf{y}$ coordinate for such a wavefunction is
undetermined:
\begin{equation}
  \label{eq:10}
  \delta \left( \frac{\mathbf{x}}{w} \right) =  \qquad   \includegraphics[valign=c]{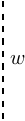}
\end{equation}
These are precisely the dashed lines that featured in~\cite{Z1,Z2,Z3}.
In this polarization $\mathbf{y}$ acts as a $\mathfrak{q}$-difference
operator $\mathfrak{q}^{\mathbf{x} \partial_{\mathbf{x}}}$, so that
$x$ coordinate of the dashed line before and after the junction with a
$(0,1)$ brane differ by $\frac{1}{R} \ln \mathfrak{q}$. We will not
attempt to draw the corresponding ``quantum'' version of the
three-dimensional diagram~\eqref{eq:7}. What we have just described is
essentially the algebra of line operators
\begin{equation}
  \label{eq:13}
  L_{\zeta}^{(n,m), U(1)} = \mathbf{x}^m
  \mathbf{y}^{-n}
\end{equation}
in $\mathcal{N}=4$ $U(1)$ gauge theory with line operators
corresponding to semi-infinite $(n,m)$ strings ending on a D3
brane. The parameter $\zeta \in U(1)$ does not enter in the relations
of the algebra, but as we will see corresponds to the overall rotation
of the picture in $\mathbb{R}^2_{xy}$ plane. It will play a prominent
role when we turn to several D3 branes and to non-abelian gauge theory
in a moment. The conceptual reason for the non-commutativity of the
algebra for $\mathfrak{q} \neq 1$ is that the line operators must sit
at the fixed point (the origin) of $\mathbb{C}_{\mathfrak{q}}$ and
therefore there is a natural ordering along $\mathbb{R}_{\tau}$.

Under the correspondence with the quantum toroidal algebra Type IIB
$S$-duality group is identified with $SL(2,\mathbb{Z})$ automorphism
group of
$U_{\mathfrak{q},\mathfrak{t}}(\widehat{\widehat{\mathfrak{gl}}}_1)$
(for trivial central charges), as detailed in
Appendix~\ref{sec:prop-repr-quant}.

The next step is to understand the algebra of line operators in $U(N)$
theory and the natural framework for this is the notion of
\emph{framed BPS states}~\cite{GMN}. A framed BPS state can be viewed
as a line operator of fixed type acting as an (infinitely) heavy probe
BPS particle, surrounded by a ``halo'' of bound BPS particles. A line
operator $L_{\zeta}$ is a UV object which at low energies (in the IR),
where the gauge group is spontaneously broken to $U(1)^N$, is expanded
in terms of line operators of each $U(1)$ factor, i.e.\ quantum torus
algebras living on each D3 brane:
\begin{equation}
  \label{eq:11}
  L_{\zeta} = \sum_{(\vec{n},\vec{m}) \in \mathbb{Z}^{2N}}
  \overline{\underline{\Omega}}(L_{\zeta}, \vec{n}, \vec{m}| \mathfrak{q},\mathfrak{t})
  \prod_{i=1}^N \mathfrak{q}^{-\frac{m_i n_i}{2}} \mathbf{x}_i^{m_i} \mathbf{y}_i^{-n_i}  
\end{equation}
where $\mathbf{x}_i$, $\mathbf{y}_i$ are generators
satisfying~\eqref{eq:16}, and $\zeta \in U(1)$ is a parameter
associated with the line operator which keeps track of the phase of
the supercharges under which $L_{\zeta}$ is invariant (we will comment
more on the role of $\zeta$ in sec.~\ref{sec:mystery-coproduct}). The
fundamental formula~\eqref{eq:11} gives a homomorphism from the
algebra of line operators (with generally unknown complicated
commutation relations) to just $N$ copies of a quantum torus. The
coefficients of the expansion have physical meaning of their own: they
are \emph{framed BPS protected spin characters} (framed PSCs for short),
counting the number of framed BPS states with given electric (resp.\
magnetic) charges $\vec{n}$ (resp.\ $\vec{m}$) under $N$ $U(1)$ gauge
groups\footnote{It is related to the protected spin character defined
  in~\cite{GMN} for $\mathcal{N}=2$ theories by $\mathfrak{q} =
  y_{\mathrm{GMN}}^2$. The parameter~$\mathfrak{t}$ is the fugacity of
  the extra $R$-symmetry appearing in $\mathcal{N}=4$ theory.}:
\begin{equation}
  \label{eq:12}
  \overline{\underline{\Omega}}(L_{\zeta}, \vec{n}, \vec{m}|\mathfrak{q},\mathfrak{t}) = \Tr_{\mathcal{H}_{\mathrm{BPS}}(L_{\zeta}, \vec{n}, \vec{m})}
  (-1)^{2J_3} \mathfrak{q}^{-J_3 - I_{3,R}} \left( \frac{\mathfrak{q}}{\mathfrak{t}^2} \right)^{I_{3,L}},
\end{equation}
where $\mathcal{H}_{\mathrm{BPS}}(L_{\zeta}, \vec{n}, \vec{m})$ is the
space of framed BPS states with line operator $L_{\zeta}$ insertion
and charges $\vec{n}$, $\vec{m}$. The operators $J_3$, $I_{3,L}$ and
$I_{3,R}$ are Cartan generators of $\mathfrak{so}(3) \simeq
\mathfrak{su}(2)$ rotations in $\mathbb{C}_{\mathfrak{q}} \times
\mathbb{R}_{\tau}$, and $\mathfrak{su}(2)_L \oplus \mathfrak{su}(2)_R
\simeq \mathfrak{so}(4)$ part of the $\mathcal{N}=4$ $R$-symmetry
respectively. These generators implement the twisted periodic boundary
conditions on the $\mathbb{C}_{\mathfrak{q}} \times
\mathbb{C}_{\mathfrak{t}^{-1}} \times
\mathbb{C}_{\mathfrak{t}/\mathfrak{q}} \times S^1$ part of the Type IIB
background from Table~\ref{tab:1}.

String theory interpretation of framed BPS states can be guessed from
our treatment of the algebra of line operators in the $U(1)$ theory
above: line operators correspond to semi-infinite strings ending on D3
branes. However, if there are several D3 branes one needs to decide on
which of them to end a given semi-infinite string. Another possibility
which arises for multiple D3 branes is a nontrivial string networks
with semi-infinite strings. In fact, as we will see in
sec.~\ref{sec:framed-bps-states}, the correct answer is a linear
combination of nontrivial string networks.

In sec.~\ref{sec:algebra-line-oper} using the interpretation of framed
BPS states in terms of string networks we demonstrate that line
operators in $\mathcal{N}=4$ $U(N)$ gauge theory correspond to
PBW-type generators $P_{(n,m)}$, $(n,m) \in \mathbb{Z}^2\backslash
(0,0)$ of the quantum toroidal algebra
$U_{\mathfrak{q},\mathfrak{t}}(\widehat{\widehat{\mathfrak{gl}}}_1)$
taken in a tensor product of $N$ vector representations
$\mathcal{V}_{\mathfrak{q}}$. The UV-IR expansion
formulas~\eqref{eq:11} for the line operators are then understood as
the $(N-1)$-fold action of the coproduct on the generators of
$U_{\mathfrak{q},\mathfrak{t}}(\widehat{\widehat{\mathfrak{gl}}}_1)$.

As one varies the parameters of the theory (vacuum moduli and the
phase $\zeta$) one encounters walls at which the
homomorphism~\eqref{eq:11} of the (UV) algebra of line operators into
the quantum torus, and hence protected spin characters
$\overline{\underline{\Omega}}(L_{\zeta}, \vec{n}, \vec{m}|q,t)$,
change discontinuously. The insight of~\cite{GMN} is that for fixed
vacuum moduli the walls encountered when varying $\arg \zeta$
correspond to standard (i.e.\ unframed) BPS states. These walls are
usually called walls of the second kind, in contrast to walls of the
first kind, which one encounters when varying vacuum moduli. The value
of $\zeta$ at a wall $W_{\mathcal{P}}$ associated with a BPS state
$\mathcal{P}$ is equal to the phase of the central charge $\arg
Z_{\mathcal{P}}$ of the BPS state.

What is the role of the phase parameter $\zeta$ in our
representation-theoretic interpretation of framed BPS states? The key
to understanding this is the fact that quantum toroidal algebra
$U_{\mathfrak{q},\mathfrak{t}}(\widehat{\widehat{\mathfrak{gl}}}_1)$
has in fact an infinite number of different coproducts, parametrized
by the choice of the Borel subalgebra which in turn depends on the
choice of a ray of \emph{irrational slope} in an $\mathbb{R}^2$
plane. In sec.~\ref{sec:mystery-coproduct} we demonstrate that the
slope parameter of the coproduct in the quantum toroidal algebra
should be identified with the phase of the parameter $\zeta$. There is
an infinite number of walls $W_{(n,m)}$ corresponding to rational
slopes $\frac{m}{n}$, each wall separating two choices of the
coproduct. Transitions between coproducts with different slopes are
implemented by a product of Drinfeld twists associated with each
wall. This structure fits in with Khoroshkin-Tolstoy formula for the
universal $R$-matrix and Kontsevich-Soibelman wall-crossing formula
for framed BPS states.

Finally, in sec.~\ref{sec:drinf-twists-konts} we explore the
possibility of combining coproducts of different slopes together and
learn that this corresponds to wall-crossing (of both the first and
the second kind) of unframed BPS states.  Conclusions and some open
problems are presented in sec.~\ref{sec:concl-open-probl}.

\section{String networks in Type IIB string theory and BPS states in gauge theory}
\label{sec:string-networks-type}

We consider $\frac{1}{4}$-BPS states in $\mathcal{N}=4$
super-Yang-Mills theory with gauge group $U(N)$. As the name suggests,
$\frac{1}{4}$-BPS states are invariant under four out of sixteen
supersymmetries of the theory. We use the conventions of~\cite{GMN}
for the $\mathcal{N}=2$ part of the supersymmetry generators and
central charges. From the point of view of $\mathcal{N}=2$
supersymmetry the states we consider are similar to those considered
in~\cite{GMN} and in~\cite{S}. In Type IIB picture
$\frac{1}{4}$-BPS states correspond to planar networks of $(p,q)$
strings formed using triple junctions with strings ending on D3
branes\footnote{In the special case when there are no triple junctions
  and the whole network consists of a single $(p,q)$ string stretched
  between a pair of D3 branes, more supersymmetry is preserved and the
  corresponding state is $\frac{1}{2}$-BPS.}.

\begin{figure}[h]
  \centering
  \includegraphics[valign=c]{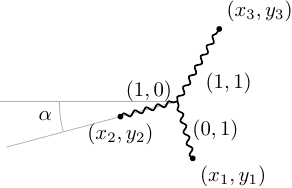}
  \caption{An example of a string network in the $\mathbb{R}^2_{xy}$
    plane consisting of three $(p,q)$ strings (drawn as wavy lines)
    stretched between three D3 branes located at points $(x_1,y_1)$,
    $(x_2,y_2)$ and $x_3, y_3$. The relative angles of the $(p,q)$
    string segments are fixed by their charges and the value of $\tau$
    (see Eq.~\eqref{eq:3}). We set $\Re \tau = 0$ in the figure, so
    that $(1,0)$ and $(0,1)$ strings are orthogonal. The overall
    angle of the network $\alpha$ is equal to the phase of the central
    charge of the corresponding $\frac{1}{4}$-BPS state.}
  \label{fig:1}
\end{figure}

Let us recall some basic properties of $(p,q)$ string
networks~\cite{S, str1, str2, str3}. The $(p,q)$ charges of the
strings must be conserved, so at any triple junction they must add up
to zero. The slopes of the strings are fixed by the BPS condition
which guarantees that the tensions at every junction are also
balanced. We denote the complex Type IIB coupling constant (which
coincides with the complex coupling constant of the $\mathcal{N}=4$
gauge theory) by $\tau \in \mathbb{C}$ ($\Im \tau >0$). Let us assume
for a moment that $\mathfrak{q}=1$ and D3 branes have definite
positions in the $\mathbb{R}^2_{xy}$ plane. To satisfy the BPS
condition a $(p,q)$ string belonging to a network $\mathcal{P}$ must
lie parallel to the vector
\begin{equation}
  \label{eq:3}
  e^{i\alpha_{\mathcal{P}}}(p \bar{\tau} + q)
\end{equation}
in the $\mathbb{R}^2_{xy}$ plane, on which we have introduced a
complex coordinate $z = x+iy$. The phase $\alpha_{\mathcal{P}}$ is
arbitrary, but the same for all strings belonging to a given network
$\mathcal{P}$; it coincides with the phase of the central charge
$Z_{\mathcal{P}}$ of the corresponding BPS state. By the definition of
BPS states, the absolute value of the central charge is equal to the
mass of the state, which in turn is the sum of masses of all edges of
the string network, each given by the product of its length $|\Delta
z|$ and $(p,q)$ string tension $T_{p,q} = \frac{1}{\sqrt{\Im \tau}} |p
\tau + q|$:
\begin{equation}
  \label{eq:4}
  |Z_{\mathcal{P}}| = \frac{1}{\sqrt{\Im \tau}} \sum_{e \in \mathrm{edges}(\mathcal{P})} |\Delta z_e| |p_e \tau
  + q_e|
\end{equation}
For given $(p,q)$ charges of the strings the phase of
$Z_{\mathcal{P}}$ (and lengths of the strings) is determined by the
positions of the D3 branes on which the strings end. The domains in
the configuration space of D3 branes in which a network with given
topology exists or not are separated by walls on which the spectrum of
$\frac{1}{4}$-BPS states jumps.

To each string network $\mathcal{P}$ one associates the space of BPS
states $\mathcal{H}_{\mathrm{BPS}}(\mathcal{P})$ (\emph{unframed,}
i.e.\ without any line operator insertion) which can be thought of as
the space of excitations of the $(p,q)$ strings with boundary
conditions given by the D3 branes. The information about
$\mathcal{H}_{\mathrm{BPS}}(\mathcal{P})$ is captured by
\emph{protected spin character} (PSC)~\cite{S} given by\footnote{Our
  notation is related to the notation of~\cite{S} by $\mathfrak{q} =
  (-y_{\mathrm{Sen}})^{-2}$, $\mathfrak{t}^{-1} = z_{\mathrm{Sen}}
  y_{\mathrm{Sen}}$.}
\begin{equation}
  \label{eq:15}
  \Omega(\mathcal{P}|\mathfrak{q},\mathfrak{t}) = -\frac{1}{\mathfrak{q}^{1/2} - \mathfrak{q}^{-1/2}} \Tr_{\mathcal{H}_{\mathrm{BPS}}(\mathcal{P})}
  (-1)^{2J_3} (2J_3) \mathfrak{q}^{-J_3 - I_{3,R}} \left( \frac{\mathfrak{q}}{\mathfrak{t}^2} \right)^{I_{3,L}}.
\end{equation}
We refer the reader to~\cite{GMN} for details about the definition of
the PSC, why it only receives contributions from BPS states and why it
is constant away from the walls. PSC~\eqref{eq:15} for unframed BPS
states plays the same role as framed PSC~\eqref{eq:12} for framed BPS
states.

Notice that $\mathfrak{q}$ is nontrivial in the definition of PSC. It
is therefore natural to ask what remains of the pictures like
Fig.~\ref{fig:1} when the non-commutativity parameter $\mathfrak{q}$
is turned on and the D3 branes become delocalized as we have discussed
in the Introduction. Naively in this case we can no longer pinpoint
the location of the strings' endpoints, and therefore it makes no
sense to talk about wall-crossing behavior of the BPS states. However,
as we will see in sec.~\ref{sec:mystery-coproduct}, the ``missing''
parameters of the configuration space for $\mathfrak{q}\neq 1$ are in
fact preserved as phase parameters of the line operators and
coproducts. In this way wall-crossing of string networks does make
sense for $\mathfrak{q}\neq 1$, although this sense is algebraic
rather than geometric. In this section, however, we simply keep the
intuitive $\mathfrak{q}=1$ picture in mind even though we will
consider PSCs with nontrivial $\mathfrak{q}$.

\begin{figure}[h]
  \centering
  \includegraphics[valign=c]{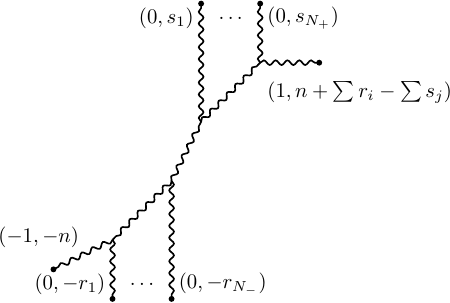}
  \caption{String network $\mathcal{P}_{\mathrm{SY}}(\vec{r},
    \vec{s})$ corresponding to a Stern-Yi bound state of dyons. $r_1,
    \ldots, r_{N_{-}}$ and $s_1, \ldots s_{N_{+}}$ are integers
    characterizing (electric) charges of the dyon under multiple
    $U(1)$ gauge groups in the IR.}
  \label{fig:2}
\end{figure}

In~\cite{S} PSC for several examples of string networks have been
computed using Kontsevich-Soibelman wall-crossing formula. We give
here some of these results since we will need them in what follows.
\begin{enumerate}
\item Let $\mathcal{P}_{(n,m)}$ denote a network consisting of a
  single $(n,m)$ string stretched between a pair of D3 branes. It
  supports a BPS state only if $n$ and $m$ are coprime. We have
  \begin{equation}
    \label{eq:17}
    \Omega (\mathcal{P}_{(n,m)}| \mathfrak{q}, \mathfrak{t}) =
    \begin{cases}
      - \left( \sqrt{\mathfrak{t}} - \frac{1}{\sqrt{\mathfrak{t}}}\right) \left( \sqrt{\frac{\mathfrak{q}}{\mathfrak{t}}} -
        \sqrt{\frac{\mathfrak{t}}{\mathfrak{q}}} \right), & \gcd(n,m) = 1,\\
      0& \text{otherwise}
    \end{cases}
  \end{equation}

\item A network $\mathcal{P}_{\mathrm{SY}}(\vec{r}, \vec{s})$ is
  depicted in Fig.~\ref{fig:2}. It corresponds to a bound state of
  dyons in super-Yang-Mills studied in by Stern and Yi~\cite{SY}. We have~\cite{S}:
  \begin{multline}
    \label{eq:18}
    \Omega(\mathcal{P}_{\mathrm{SY}}(\vec{r},
    \vec{s})|\mathfrak{q},\mathfrak{t}) =\\
    =-\left(\left( \sqrt{\mathfrak{t}} - \frac{1}{\sqrt{\mathfrak{t}}}\right) \left( \sqrt{\frac{\mathfrak{q}}{\mathfrak{t}}} -
        \sqrt{\frac{\mathfrak{t}}{\mathfrak{q}}} \right) \right)^{N_{+} + N_{-} + 1} \prod_{i=1}^{N_{+}} \frac{\mathfrak{q}^{r_i/2} -
      \mathfrak{q}^{-r_i/2}}{\mathfrak{q}^{1/2} - \mathfrak{q}^{-1/2}} \prod_{j=1}^{N_{-}} \frac{\mathfrak{q}^{s_j/2} -
      \mathfrak{q}^{-s_j/2}}{\mathfrak{q}^{1/2} - \mathfrak{q}^{-1/2}}
  \end{multline}

  \end{enumerate}

\subsection{Framed BPS states from string networks}
\label{sec:framed-bps-states}
In this section we use Sen's results for PSCs of unframed BPS states
to guess PSCs for framed BPS states with insertions of the simplest
possible line operators, i.e.\ the Wilson line in the fundamental
representation $\mathbb{C}^N$ of $U(N)$ and charge one 't Hooft line.

The line operators featuring in the framed BPS states can be thought
of as very heavy BPS particles. Since the mass of a string segment of
a network is proportional to its length, a heavy BPS particle
corresponds to a very long string. We can think of this long string as
ending on a ``probe'' D3 brane very far away. In the limit of
infinitely heavy particle the string becomes semi-infinite and the
probe D3 brane is sent to infinity. This fits nicely with the picture
of line operators in $U(1)$ theory as semi-infinite strings, e.g.\
Eq.~\eqref{eq:6}.

The probe D3 brane that is sent to infinity naively disappears from
the picture. However, information about the direction along which it
is sent to infinity is actually retained in the form of the phase
parameter $\zeta$. Indeed, as follows from Eq.~\eqref{eq:3}, for fixed
$(p,q)$ charges of the string its slope in the picture determines the
phase of the central charge of the BPS state corresponding to the
string network. This precisely reproduces the definition of the phase
parameter $\zeta$ of a framed BPS state: it is essentially the phase
of the central charge of the heavy BPS particle serving as a ``core''
of the framed BPS state.

Summarizing, we find that framed BPS states are string networks with
semi-infinite strings. If a network contains semi-infinite $(p,q)$
string at angle $\alpha_{p,q}$ in the $\mathbb{R}^2_{x,y}$ plane, then the
argument of the phase parameter $\zeta$ of the line operator in the
``core'' of the framed BPS state is given by
\begin{equation}
  \label{eq:19}
  \arg \zeta = \alpha_{p,q} - \arg (p \bar{\tau} +
  q).
\end{equation}
If there happens to be several semi-infinite strings in a given
network, the definition~\eqref{eq:19} gives the same $\arg \zeta$ for
any of them due to the BPS condition~\eqref{eq:3}.

The unframed PSC for a network is locally independent of the lengths
of the $(p,q)$ string segments. Sending a D3 brane to infinity should
not affect the value of the PSC of the network, as long as it retains
the same topology and does not cross any walls. We can thus, use the
results of~\cite{S} to get some framed PSCs, and most importantly, to
guess the expression of the form~\eqref{eq:11} for the line operators.

Let us begin with the simplest nontrivial case of $N=2$ D3
branes. Suppose $\arg \zeta = -\frac{\pi}{2}$ and we consider a
semi-infinite $(0,1)$ string which physically should correspond to a
Wilson line $L_{e^{-\frac{i\pi}{2}}}^{(0,1)}$ in $\mathbb{C}^2$
representation of the $U(2)$ gauge group. Since the ``probe'' D3 brane
is infinitely far away, the $(0,1)$ string can end either on the first
or the second D3 brane and still satisfy the BPS condition. We denote
these two possibilities by
\begin{equation}
  \label{eq:21}
  \mathcal{P}_{(0,1)}^{(1)} 
  =\qquad  \includegraphics[valign=c]{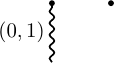}\qquad
  \qquad \qquad
  \mathcal{P}_{(0,1)}^{(2)} 
  =\qquad   \includegraphics[valign=c]{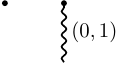}
\end{equation}
From Eq.~\eqref{eq:17} we get
\begin{align}
  \label{eq:2}
  \Omega(\mathcal{P}_{(0,1)}^{(1)}| \mathfrak{q},\mathfrak{t}) =
  \Omega(\mathcal{P}_{(0,1)}^{(2)}| \mathfrak{q},\mathfrak{t}) = -
  \left( \sqrt{\mathfrak{t}} - \frac{1}{\sqrt{\mathfrak{t}}}\right)
  \left( \sqrt{\frac{\mathfrak{q}}{\mathfrak{t}}} -
    \sqrt{\frac{\mathfrak{t}}{\mathfrak{q}}} \right).
\end{align}
We assume that the D3 branes sit on the same horizontal line. Then it
is not possible to have a string network with a triple junction and a
single semi-infinite $(0,1)$ string and therefore the two networks in
Eq.~\eqref{eq:21} are the only possible terms in the UV-IR expansion
of $L_{e^{-\frac{i\pi}{2}}}^{(0,1)}$. Physics perspective suggests
that the line operator $L_{e^{-\frac{i\pi}{2}}}^{(0,1)}$ should
decompose similarly to the weight decomposition of the $\mathbb{C}^2$
representation of $U(2)$, which corresponds to the sum of the networks
$\mathcal{P}_{(0,1)}^{(1)}$ and $\mathcal{P}_{(0,1)}^{(2)}$ (the
endpoints of the strings contribute the quantum torus operators
$\mathbf{x}_i$, $\mathbf{y}_i$). Our guess for the UV-IR expansion of
the Wilson line is therefore
\begin{equation}
  \label{eq:22}
  L_{e^{-\frac{i\pi}{2}}}^{(0,1),U(2)} = \mathcal{P}_{(0,1)}^{(1)} +
  \mathcal{P}_{(0,1)}^{(2)} = - \left( \sqrt{\mathfrak{t}} - \frac{1}{\sqrt{\mathfrak{t}}}\right) \left( \sqrt{\frac{\mathfrak{q}}{\mathfrak{t}}} -
        \sqrt{\frac{\mathfrak{t}}{\mathfrak{q}}} \right) (\mathbf{x}_1
      + \mathbf{x}_2).
\end{equation}

Next we consider line a operator $L_{e^{-\frac{i\pi}{2}}}^{(1,n)}$
presumably corresponding to a Wilson-'t Hooft line. The semi-infinite
$(1,n)$ string can still join the first or the second D3 brane giving
rise to two contributions similar to Eq.~\eqref{eq:21}:
\begin{equation}
  \label{eq:20}
\mathcal{P}_{(1,n)}^{(1)}
=\qquad  \includegraphics[valign=c]{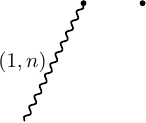} \qquad
\qquad \qquad \mathcal{P}_{(1,n)}^{(2)} = \qquad
\includegraphics[valign=c]{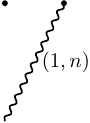}  
\end{equation}
However, in this case there are other possibilities involving a triple
junction which we denote by $\tilde{\mathcal{P}}^{(k)}_{(1,n)}$:
\begin{equation}
  \label{eq:23}
  \tilde{\mathcal{P}}^{(k)}_{(1,n)} = \qquad  \includegraphics[valign=c]{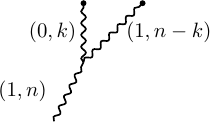}
\end{equation}
The PSCs of the networks~\eqref{eq:20}~\eqref{eq:23} can be found from
Eqs.~\eqref{eq:17},~\eqref{eq:18} and we get
\begin{align}
  \label{eq:24}
  \Omega(\mathcal{P}_{(1,n)}^{(1)}| \mathfrak{q},\mathfrak{t}) &=
  \Omega(\mathcal{P}_{(1,n)}^{(2)}| \mathfrak{q},\mathfrak{t}) = -
  \left( \sqrt{\mathfrak{t}} - \frac{1}{\sqrt{\mathfrak{t}}}\right)
  \left( \sqrt{\frac{\mathfrak{q}}{\mathfrak{t}}} -
    \sqrt{\frac{\mathfrak{t}}{\mathfrak{q}}} \right),\\
  \Omega(\tilde{\mathcal{P}}_{(1,n)}^{(k)}| \mathfrak{q},\mathfrak{t})
  &= -\left(\left( \sqrt{\mathfrak{t}} -
      \frac{1}{\sqrt{\mathfrak{t}}}\right) \left(
      \sqrt{\frac{\mathfrak{q}}{\mathfrak{t}}} -
      \sqrt{\frac{\mathfrak{t}}{\mathfrak{q}}} \right) \right)^2
  \frac{\mathfrak{q}^{k/2} - \mathfrak{q}^{-k/2}}{\mathfrak{q}^{1/2} -
    \mathfrak{q}^{-1/2}}
\end{align}
We notice that the contributions of the networks
$\tilde{\mathcal{P}}_{(1,n)}^{(k)}$ taken with the corresponding
quantum torus operators $\mathbf{x}_i$, $\mathbf{y}_i$ can be packed
into a very nice generating function:
\begin{equation}
  \label{eq:25}
  \sum_{k \geq 1} \Omega(\tilde{\mathcal{P}}_{(1,n)}^{(k)}|
  \mathfrak{q},\mathfrak{t}) \mathbf{x}_1^k \mathfrak{q}^{\frac{k-n}{2}} \mathbf{x}_2^{n-k}
  \mathbf{y}_2^{-1} = - \left( \sqrt{\mathfrak{t}} -
    \frac{1}{\sqrt{\mathfrak{t}}}\right) \left(
    \sqrt{\frac{\mathfrak{q}}{\mathfrak{t}}} -
    \sqrt{\frac{\mathfrak{t}}{\mathfrak{q}}} \right)  \left(
    \frac{\left( 1 - \frac{\mathfrak{q}}{\mathfrak{t}} \frac{\mathbf{x}_1}{\mathbf{x}_2}
      \right) \left( 1 - \mathfrak{t} \frac{\mathbf{x}_1}{\mathbf{x}_2}
      \right)}{\left( 1 -  \mathfrak{q} \frac{\mathbf{x}_1}{\mathbf{x}_2}
      \right)\left( 1 -  \frac{\mathbf{x}_1}{\mathbf{x}_2}
      \right)} - 1 \right) \mathfrak{q}^{-\frac{n}{2}} \mathbf{x}_2^n \mathbf{y}_2^{-1}.
\end{equation}
This suggests that we need to take a sum of all three types of
networks: $\mathcal{P}_{(1,n)}^{(1)}$,
$\mathcal{P}_{(1,n)}^{(2)}$ and
$\tilde{\mathcal{P}}_{(1,n)}^{(k)}$ together to get the line operator
$L_{e^{-\frac{i\pi}{2}}}^{(1,n)}$:
\begin{multline}
  \label{eq:26}
  L_{e^{-\frac{i\pi}{2}}}^{(1,n),U(2)} = \mathcal{P}_{(1,n)}^{(1)} +
  \mathcal{P}_{(1,n)}^{(2)}+ \sum_{k \geq 1}
  \tilde{\mathcal{P}}_{(1,n)}^{(k)} =\\
  = - \left( \sqrt{\mathfrak{t}} -
      \frac{1}{\sqrt{\mathfrak{t}}}\right) \left(
      \sqrt{\frac{\mathfrak{q}}{\mathfrak{t}}} -
      \sqrt{\frac{\mathfrak{t}}{\mathfrak{q}}} \right) \mathfrak{q}^{-\frac{n}{2}} \left(
      \mathbf{x}_1^n \mathbf{y}_1^{-1} + 
    \frac{\left( 1 - \frac{\mathfrak{q}}{\mathfrak{t}} \frac{\mathbf{x}_1}{\mathbf{x}_2}
      \right) \left( 1 - \mathfrak{t} \frac{\mathbf{x}_1}{\mathbf{x}_2}
      \right)}{\left( 1 -  \mathfrak{q} \frac{\mathbf{x}_1}{\mathbf{x}_2}
      \right)\left( 1 -  \frac{\mathbf{x}_1}{\mathbf{x}_2}
      \right)}   \mathbf{x}_2^n \mathbf{y}_2^{-1} \right)
\end{multline}

Remarkably, our heuristic derivation has produced a family of line
operators $L_{e^{-\frac{i\pi}{2}}}^{(1,n)}$ which coincide with the
generators $P_{(1,n)} \in
U_{\mathfrak{q},\mathfrak{t}}(\widehat{\widehat{\mathfrak{gl}}}_1)$
taken in a tensor product $\mathcal{V}_{\mathfrak{q}}^{*}
\otimes_{-\frac{\pi}{2}-\epsilon} \mathcal{V}_{\mathfrak{q}}^{*}$ of
two vector representations\footnote{These operators are also related
  by conjugation to trigonometric Ruijsenaars-Schneider (also known as
  Macdonald) difference operators.} from Eq.~\eqref{eq:48}. More
details on vector representations and their tensor products are
presented in sec.~\ref{sec:vect-repr}. The contribution of nontrivial
string networks $\tilde{\mathcal{P}}_{(1,n)}^{(k)}$ appears due to the
nontriviality of the coproduct on the quantum toroidal algebra
$U_{\mathfrak{q},\mathfrak{t}}(\widehat{\widehat{\mathfrak{gl}}}_1)$. The
line operator $L_{e^{-\frac{i\pi}{2}}}^{(0,1)}$ also matches with the
action of the corresponding generators $P_{(0,1)}$ in the tensor
product of vector representations, as prescribed by
Eq.~\eqref{eq:14}. Similar calculation can be done for
$L_{e^{-\frac{i\pi}{2}}}^{(-1,n)}$ matching Eq.~\eqref{eq:5}.

Of course, the arguments that we have given for the
expansion~\eqref{eq:26} are to a large extent based on guesswork. It
is important to get a more rigorous and general calculation of
possible string networks contributing to a given line operator, but we
leave this for future work.

\subsection{Noncommutative algebra of line operators}
\label{sec:algebra-line-oper}
Given the known relation of D3 branes with vector representations of
quantum toroidal algebra
$U_{\mathfrak{q},\mathfrak{t}}(\widehat{\widehat{\mathfrak{gl}}}_1)$
explained in sec.~\ref{sec:introduction} and the calculations in
sec.~\ref{sec:framed-bps-states}, we can try to guess the full algebra
of line operators in $\mathcal{N}=4$ $U(N)$ theory. Indeed, the
elements $P_{(\pm 1,n)}$ generate the full quantum toroidal algebra by
successive commutators. Therefore, we expect that the algebra of line
operators is the quantum toroidal algebra acting in a tensor product
of $N$ vector representations with generators
$\rho_{(\mathcal{V}^{*}_{\mathfrak{q}})^{\otimes N}}(P_{(n,m)})$. In
sec.~\ref{sec:mystery-coproduct} we will see that the tensor product
requires a choice of coproduct, which leads to different realizations
of the line operator algebra related by wall-crossing transitions,
which we relate to Drinfeld twists.

Another conjectured description of the algebra of line operators in
$U(N)$ theory is the spherical double affine Hecke algebra (sDAHA)
$\mathbb{SH}_{\mathfrak{q},\mathfrak{t}}(N)$~\cite{V}. It is,
however, equivalent to the quotient of the quantum toroidal algebra
that we have just described. Indeed, the tensor product of $N$ vector
representations of
$U_{\mathfrak{q},\mathfrak{t}}(\widehat{\widehat{\mathfrak{gl}}}_1)$
is known to be the same as the faithful representation of
$\mathbb{SH}_{\mathfrak{q},\mathfrak{t}}(N)$ by
$\mathfrak{q}$-difference operators in $N$ variables.

For gauge theories with simple gauge groups $SU(2)$ and $SO(3)$ (with
possible discrete theta-angle) the algebra of line operators has been
studied in~\cite{GMN, GK} with similar conclusions: they
are quotients of $\mathbb{SH}_{\mathfrak{q},\mathfrak{t}}(2)$ by a
combination of two $\mathbb{Z}_2$ involutions $\sigma_1$ and
$\sigma_2$:
\begin{equation}
  \label{eq:27}
  P_{(n,m)} \stackrel{\sigma_1}{\leftrightarrow} (-1)^n P_{(n,m)},
  \qquad \qquad   P_{(n,m)} \stackrel{\sigma_2}{\leftrightarrow} (-1)^m P_{(n,m)}.
\end{equation}

\section{Choices of coproducts}
\label{sec:mystery-coproduct}
In sec.~\ref{sec:string-networks-type} we have understood the algebra
of line operators in $U(N)$ gauge theory using string
networks. However, we have not explained the wall-crossing behaviour
of the framed BPS states as the parameters of the theory are
varied. There are two types of parameters in the theory: the phase
parameter $\zeta$ associated with a line operator $L_{\zeta}$ and
vacuum moduli of the gauge theory. We will deal with them in turn.  As
we have mentioned in sec.~\ref{sec:framed-bps-states} the phase
parameter $\zeta$ determines the overall angle at which the
semi-infinite $(p,q)$ strings arrive into the picture. Although our
picture with localized D3 branes is a semiclassical approximation
valid only in the limit $\mathfrak{q}\to 1$, we can still expect from
it that if one varies the angle at which a $(0,1)$ string arrives into
a system of D3 branes, string networks of different topology become
possible.

To get an algebraic interpretation of $\zeta$ it is instructive to
analyze the action of $SL(2,\mathbb{Z})$ duality on it. In fact this
analysis has been done in~\cite{KW} (sec.~3.1), where it was found
that under the action of an element $\left(
  \begin{smallmatrix}
    a& b\\
    c& d
  \end{smallmatrix}
\right) \in SL(2,\mathbb{Z})$ the central charges transform so that
\begin{equation}
  \label{eq:28}
  \zeta \to \frac{|c\tau + d|}{c \tau + d} \zeta
\end{equation}
or more explicitly
\begin{equation}
  \label{eq:29}
  \zeta \stackrel{\mathcal{T}}{\to} \zeta, \qquad \qquad \zeta
  \stackrel{\mathcal{S}}{\to} -\frac{|\tau|}{\tau} \zeta
\end{equation}
where $\mathcal{S}$ and $\mathcal{T}$ are the standard $SL(2,
\mathbb{Z})$ generators. For $\Re \tau = 0$ the phase $\zeta$ is
invariant under $\mathcal{T}$ and is rotated by $e^{i\frac{\pi}{2}}$
under $\mathcal{S}$\footnote{For $\Re \tau \neq 0$ one needs to make a
  simple reparametrization.}. This coincides with the transformation
law of the phase determining the Borel subalgebra $\mathcal{B}_{\arg
  \zeta}$ and hence the coproduct $\Delta_{\arg \zeta}$ on
$U_{\mathfrak{q},\mathfrak{t}}(\widehat{\widehat{\mathfrak{gl}}}_1)$
(see the details in sec.~\ref{sec:coproducts}). We can, therefore,
provisionally identify the phase parameter $\zeta$ of the line
operator $L_{\zeta}$ and the phase determining the coproduct
$\Delta_{\arg \zeta}$ on
$U_{\mathfrak{q},\mathfrak{t}}(\widehat{\widehat{\mathfrak{gl}}}_1)$.

The phase of the coproduct of the quantum toroidal algebra must have
irrational slope, so all rational slopes can be thought of as (a dense
set of) walls on the circle. The walls on the $\zeta$-circle in gauge
theory on which the spectrum of framed BPS states has discontinuities
correspond to the arguments of the central charges of the unframed BPS
states~\cite{GMN}. Let us look at these unframed BPS states for the
case of a pair of D3 branes. The only BPS states for $U(2)$ theory are
$\frac{1}{2}$-BPS dyons corresponding to a single $(n,m)$ string
stretched between two D3 branes with $\gcd(n,m)=1$. For D3 branes
located on the $x$ axis in $\mathbb{R}^2_{xy}$ (as e.g.\ in
Eq.~\eqref{eq:21}) these states have phases of the central charges
given by
\begin{equation}
  \label{eq:30}
  \phi_{n,m} = \arg Z = -\arg(n \bar{\tau} + m) = \arg(n \tau
+ m)
\end{equation}
For $\Re \tau = 0$ we have simply $\phi_{n,m} = \mathrm{Arctg}
\frac{n}{m}$. We conclude that the phases $\phi_{n,m}$ are the walls
$W_{n,m}$ on the $\zeta$ circle where framed BPS counts jump, and they
coincide with the ``forbidden'' values of the phase of the coproduct
$\Delta_{\arg \zeta}$.

According to the general theory~\cite{GMN} when going through a wall
$W_{n,m}$ on the $\zeta$ circle line operator $L^{(n,m)}_{\zeta}$ is
conjugated by an operator $S(W_{n,m})$ of the form
\begin{equation}
  \label{eq:50}
 S(W_{n,m}) = \exp \left[\sum_{k \geq 1} \frac{1}{k(\mathfrak{q}^{k/2}-\mathfrak{q}^{-k/2})} \Omega(\mathcal{P}_{n,m}|\mathfrak{q}^k,
    \mathfrak{t}^k)
    \left(\frac{\mathbf{x}_1}{\mathbf{x}_2}\right)^{k n}  \left(\frac{\mathbf{y}_1}{\mathbf{y}_2}\right)^{-k m} \right]
\end{equation}
where $\mathbf{x}_i$, $\mathbf{y}_i$ are quantum torus generators
living on the two D3 branes, $\mathcal{P}_{(n,m)}$ is the network
consisting of a single $(n,m)$ string between two D3 branes and
$\Omega(\mathcal{P}_{(n,m)}|\mathfrak{q} \mathfrak{t})$ is its PSC
given by Eq.~\eqref{eq:17}. Plugging Eq.~\eqref{eq:17} into
Eq.~\eqref{eq:50} we find
\begin{equation}
  \label{eq:51}
  S(W_{n,m}) =\exp \left[\sum_{k \geq 1} \frac{\kappa_k}{k(\mathfrak{q}^{k/2}-\mathfrak{q}^{-k/2})^2}
    \left(\frac{\mathbf{x}_1}{\mathbf{x}_2}\right)^{k n}
    \left(\frac{\mathbf{y}_1}{\mathbf{y}_2}\right)^{-k m} \right] =
  \Phi \left( \left(\frac{\mathbf{x}_1}{\mathbf{x}_2}\right)^n
    \left(\frac{\mathbf{y}_1}{\mathbf{y}_2}\right)^{-m} \right),
\end{equation}
where $\kappa_k = (1-\mathfrak{q}^k) (1-\mathfrak{t}^{-k})(1 -
\mathfrak{t}^k/\mathfrak{q}^k)$ and $\Phi(x) =
\frac{(x;\mathfrak{q})_{\infty}
  (\mathfrak{q} x;\mathfrak{q})_{\infty}}{(\mathfrak{t}
  x;\mathfrak{q})_{\infty} ( \mathfrak{q} x/\mathfrak{t} ;\mathfrak{q})_{\infty}}$. We can finally notice that
Eq.~\eqref{eq:51} coincides with the ``elementary Drinfeld twist''
$F_{\mathrm{Arctg} \frac{n}{m}}\in
U_{\mathfrak{q},\mathfrak{t}}(\widehat{\widehat{\mathfrak{gl}}}_1)
\hat{\otimes}
U_{\mathfrak{q},\mathfrak{t}}(\widehat{\widehat{\mathfrak{gl}}}_1)$
given by Eq.~\eqref{eq:42} evaluated in a pair of vector
representations using Eq.~\eqref{eq:47}:
\begin{equation}
  \label{eq:55}
  S(W_{n,m}) = \rho_{\mathcal{V}^{*}_{\mathfrak{q}}} \otimes
  \rho_{\mathcal{V}^{*}_{\mathfrak{q}}} (F_{\mathrm{Arctg} \frac{n}{m}}).
\end{equation}
The
Drinfeld twist $F_{\mathrm{Arctg} \frac{n}{m}}$ transforms the coproduct
$\Delta_{\mathrm{Arctg} \frac{n}{m} - \epsilon}$ on one side of the
wall $W_{n,m}$ into the coproduct $\Delta_{\mathrm{Arctg} \frac{n}{m}
  + \epsilon}$ on the other side of the wall:
\begin{equation}
  \label{eq:52}
  \Delta_{\mathrm{Arctg} \frac{n}{m}
    + \epsilon}(g) = F_{\mathrm{Arctg} \frac{n}{m}} \Delta_{\mathrm{Arctg} \frac{n}{m}
    - \epsilon}(g) F_{\mathrm{Arctg} \frac{n}{m}}^{-1}.
\end{equation}
for any $g \in
U_{\mathfrak{q},\mathfrak{t}}(\widehat{\widehat{\mathfrak{gl}}}_1)$.

The following consistent dictionary between the gauge theory and
representation theory arises. Line operators $L^{(n,m)}_{\zeta}$ in
$U(2)$ theory are given by the generators $P_{(n,m)}$ of the quantum
toroidal algebra evaluated in a tensor product of vector
representations using a coproduct $\Delta_{\arg \zeta}$, i.e.\
\begin{equation}
  \label{eq:53}
  L^{(n,m), U(2)}_{\zeta} = \rho_{\mathcal{V}^{*}_{\mathfrak{q}}} \otimes \rho_{\mathcal{V}^{*}_{\mathfrak{q}}} ( \Delta_{\arg
    \zeta} (P_{(n,m)})).
\end{equation}
The parameter $\zeta$ must have irrational slope. Changing the phase
of $\zeta$ by an infinitesimal amount from $\mathrm{Arctg} \frac{k}{l}
- \epsilon$ to $\mathrm{Arctg} \frac{k}{l} + \epsilon$ means crossing
a wall $W_{k,l}$ of rational slope $\frac{k}{l}$ and such a crossing
changes the coproduct by an elementary Drinfeld twist corresponding to
the wall $W_{k,l}$:
\begin{equation}
  \label{eq:58}
  L^{(n,m), U(2)}_{\exp \left( i \mathrm{Arctg} \frac{k}{l} + i\epsilon
    \right)} =
  S(W_{k,l})^{-1} L^{(n,m), U(2)}_{\exp \left( i \mathrm{Arctg} \frac{k}{l} 
    -i\epsilon \right)} S(W_{k,l}). 
\end{equation}
Changing the phase by a finite amount means crossing an
infinite number of walls, and is implemented by the conjugation with
``macroscopic'' Drinfeld twist $F_{\vartheta, \vartheta'}$ given by
Eq.~\eqref{eq:41}:
\begin{equation}
  \label{eq:59}
  L^{(n,m), U(2)}_{\exp (i \vartheta')} =
 \rho_{\mathcal{V}^{*}_{\mathfrak{q}}} \otimes
 \rho_{\mathcal{V}^{*}_{\mathfrak{q}}} (F_{\vartheta, \vartheta'}^{-1}) L^{(n,m), U(2)}_{\exp (i \vartheta)} \rho_{\mathcal{V}^{*}_{\mathfrak{q}}} \otimes
 \rho_{\mathcal{V}^{*}_{\mathfrak{q}}} (F_{\vartheta, \vartheta'}). 
\end{equation}

A particularly interesting case of wall-crossing is the rotation of
the phase of $\zeta$ by $\pi$. The corresponding operator is nothing
but the $R$-matrix of the quantum toroidal algebra (evaluated in a
pair of vector representations). Indeed, one can notice from
Eq.~\eqref{eq:44} that $\rho_{\mathcal{V}^{*}_{\mathfrak{q}}} \otimes
\rho_{\mathcal{V}^{*}_{\mathfrak{q}}} \Delta_{\arg \zeta + \pi} =
\rho_{\mathcal{V}^{*}_{\mathfrak{q}}} \otimes
\rho_{\mathcal{V}^{*}_{\mathfrak{q}}} \Delta_{\arg
  \zeta}^{\mathrm{op}}$. Moreover the identification with
wall-crossing automatically reproduces the factorized
Khoroshkin-Tolstoy form of the $R$-matrix~\cite{KT1, KT2}. On the
gauge theory side the product over all phases of the central charges
reproduces the Kontsevich-Soibelman spectrum generator~\cite{KS}:
\begin{equation}
  \label{eq:60}
  \rho_{\mathcal{V}^{*}_{\mathfrak{q}}} \otimes
\rho_{\mathcal{V}^{*}_{\mathfrak{q}}} (\mathcal{R}_{\vartheta}) =
\prod^{\to}_{
  \begin{smallmatrix}
    n,m\\
    gcd(n,m)=1\\
    \vartheta < \mathrm{Arctg} \frac{n}{m} \leq \vartheta + \pi
  \end{smallmatrix}
} S(W_{n,m}),
\end{equation}
where a wall $W_{n,m}$ corresponds to the $\frac{1}{2}$-BPS
$(n,m)$-dyon in the low-energy effective $U(1)^2$ theory arising from
the UV $U(2)$ gauge theory.

This is a nice picture, however, it still does not capture all aspects
of wall-crossing in $\mathcal{N}=4$ gauge theory. Indeed, the unframed
BPS spectrum in $U(2)$ theory consists of only $\frac{1}{2}$-BPS
particles, which don't undergo any wall-crossing at all for generic
$\tau$ and are stable everywhere in the vacuum moduli space (i.e.\ for
all positions of D3 branes).  For $N \geq 3$ this is no longer the
case: $\frac{1}{4}$-BPS states arise and do undergo nontrivial
wall-crossing of the first kind as the vacuum moduli are varied. In
sec.~\ref{sec:drinf-twists-konts} we sketch how the general $U(N)$
case should work.

\subsection{Combining multiple coproducts}
\label{sec:drinf-twists-konts}
Before describing the higher rank gauge groups, let us make a short
remark about the geometric interpretation of the phase parameter in
the coproduct.  Eq.~\eqref{eq:19} implies that the physically relevant
parameter is the relative angle between the finite $(p,q)$ strings in
the network (determined by the positions of D3 branes) and the phase
parameter associated with semi-infinite strings. Throughout
sec.~\ref{sec:mystery-coproduct} we have assumed that the pair of D3
branes that we consider lies fixed on the horizontal axis in the
$\mathbb{R}_{xy}^2$ plane. If the D3 branes are rotated by $\alpha$
(say around the origin), the phase entering the coproduct in
Eq.~\eqref{eq:53} will shift by $\alpha$ too. In this way one can
either rotate a pair D3 branes keeping the phase of the $\zeta$
parameter fixed or vice versa and encounter the same walls.

It is not hard to guess what happens to our algebraic picture in the
case of more than two D3 branes. Indeed, we expect an action of
$U_{\mathfrak{q},\mathfrak{t}}(\widehat{\widehat{\mathfrak{gl}}}_1)$
on a tensor product of $N$ vector representations, each corresponding
to a D3 brane. However, now we have more choices of coproducts to
make. Indeed, the analogue of Eq.~\eqref{eq:53} for three vector
representations would be
\begin{equation}
  \label{eq:54}
  L^{(n,m), U(3)}_{\zeta} = \rho_{\mathcal{V}^{*}_{\mathfrak{q}}} \otimes
  \rho_{\mathcal{V}^{*}_{\mathfrak{q}}} \otimes
  \rho_{\mathcal{V}^{*}_{\mathfrak{q}}} ( (\Delta_{\vartheta} \otimes 1)
  (\Delta_{\vartheta'} (P_{(n,m)})))
\end{equation}
Notice that the phases $\vartheta$, $\vartheta'$ of the coproducts are
arbitrary. The composition $(\Delta_{\vartheta} \otimes 1)
\Delta_{\vartheta'}$ of coproducts with different slopes is still
compatible with multiplication in the quantum toroidal algebra, i.e.\
the map $\rho_{\mathcal{V}^{*}_{\mathfrak{q}}} \otimes
\rho_{\mathcal{V}^{*}_{\mathfrak{q}}} \otimes
\rho_{\mathcal{V}^{*}_{\mathfrak{q}}} (\Delta_{\vartheta} \otimes 1)
\Delta_{\vartheta'}$ gives a homomorphism from
$U_{\mathfrak{q},\mathfrak{t}}(\widehat{\widehat{\mathfrak{gl}}}_1)$
to the direct sum of three quantum tori living on three D3 branes. We
conjecture that the additional parameters $\vartheta$, $\vartheta'$ in
Eq.~\eqref{eq:54} correspond to the angles between the lines
connecting pairs of nearby D3 branes. 

This gives the tentative answer for the paradox we have encountered in
sec.~\ref{sec:introduction}: for generic $\mathfrak{q}$ the D3 branes
are delocalized in $\mathbb{R}_{xy}^2$ plane because of the
noncommutativity of $x$ and $y$, yet we need their precise positions
to determine whether a string network stretched between them is stable
or not. We propose that instead of fixing the coordinates of the D3
branes in the $\mathfrak{q} \neq 1$ case it is enough to fix relative
angles between the neighbouring pairs of branes, and that these angles
are responsible for the wall-crossing of both framed and unframed BPS
states. Varying the relative angles between pairs of D3 branes
one should encounter walls of the first kind, while rotating the whole
system of D3 branes will produce walls of the second kind.

Notice that for coproducts with different phase parameters
coassoiativity in general is not expected to hold:
\begin{equation}
  \label{eq:56}
  (\Delta_{\vartheta} \otimes 1 ) \Delta_{\vartheta'} (g) \neq   (1
  \otimes \Delta_{\vartheta} ) \Delta_{\vartheta'} (g),
\end{equation}
where $g \in
U_{\mathfrak{q},\mathfrak{t}}(\widehat{\widehat{\mathfrak{gl}}}_1)$ is
arbitrary.  However, an analogue of coassociativity involving Drinfeld
twists holds:
\begin{equation}
  \label{eq:57}
  (\Delta_{\vartheta} \otimes 1 ) \Delta_{\vartheta'} (g)   =
  (F_{\vartheta \vartheta'} \otimes 1) (1 \otimes F_{\vartheta, \vartheta'}^{-1})  (1
  \otimes \Delta_{\vartheta} ) \Delta_{\vartheta'} (g) (1 \otimes F_{\vartheta, \vartheta'})  (F_{\vartheta,
    \vartheta'}^{-1} \otimes 1) .
\end{equation}

Thus, the change of the \emph{relative} angles between coproducts
should reproduce the wall-crossing formulas for unframed
$\frac{1}{4}$-BPS particles. We leave the details of this for future
work.

\section{Conclusions and outlook}
\label{sec:concl-open-probl}
We have used the connection between Type IIB string networks and BPS
states in four-dimensional $\mathcal{N}=4$ $U(N)$ super Yang-Mills
theory to elucidate the structure of the algebra of line operators in
the gauge theory. We have found that the algebra is a certain quotient
of the quantum toroidal algebra
$U_{\mathfrak{q},\mathfrak{t}}(\widehat{\widehat{\mathfrak{gl}}}_1)$
which can also be understood as the sDAHA
$\mathbb{SH}_{\mathfrak{q},\mathfrak{t}}(N)$. We have understood the
UV-IR map for the line operators as a coproduct acting on
$U_{\mathfrak{q},\mathfrak{t}}(\widehat{\widehat{\mathfrak{gl}}}_1)$
generators and have identified wall-crossing of framed BPS states with
the Drinfeld twist of the coproduct. This provided a natural map
between Kontsevich-Soibelman operator and Khoroshkin-Tolstoy form of
the universal $R$-matrix for
$U_{\mathfrak{q},\mathfrak{t}}(\widehat{\widehat{\mathfrak{gl}}}_1)$.

There are several ways to develop of our results further. The
dictionary between quantum toroidal algebras and Type IIB string theor
is very general, so we can include for example D3 branes lying in
$\mathbb{C}_{\mathfrak{t}^{-1}}$ or
$\mathbb{C}_{\mathfrak{t}/\mathfrak{q}}$ planes, giving $4d$ theories
interacting along codimension two defects. Another natural possibility
is to include 5-branes. We expect a lot of interesting results in this
direction.

A more radical generalization of the setup that we have considered is
to open up the compactification circle $S^1$. This would correspond to
promoting all PSCs to actual spaces of BPS states and thus to
categorifying the algebraic structure that we have described.

Let us mention how our results fit into the framework
of~\cite{V}. There, the authors have considered the algebra of
monopole operators in $3d$ $\mathcal{N}=4$ gauge theories. These
algebras turned out to be given by representations of shifted Yangians
by difference operators. Our setup can be viewed as an uplift of this
picture to a $4d$ theory with and extra adjoint multiplet (and hence
twice as much supersymmetry) compactified on a circle of finite
radius. This leads to the generalization of Yangians in two ways: an
extra circle promotes them to quantum affine algebras, while an extra
adjoint field gives a further affinization of a quantum affine algebra
arriving at the quantum toroidal algebra
$U_{\mathfrak{q},\mathfrak{t}}(\widehat{\widehat{\mathfrak{gl}}}_1)$.

It would also be very interesting to interpret the results presented
above on the AdS side of the holographic duality.

\section*{Acknowledgements}
\label{sec:acknowledgements}
This work is supported by the European Research Council under the
European Union's Horizon 2020 research and innovation programme under
grant agreement No~948885 and by the Royal Society University Research
Fellowship.

The author would like to thank Jean-Emile Bourgine, Tudor Dimofte,
Dmitry Galakhov, Can Koz\c{c}az, Alexei Latyntsev and Sasha Shapiro
for discussions. Part of this work was done during the
``Representations, Moduli and Duality'' program at the Bernoulli
Centre at EPFL and the ``Chiralization and QFT'' workshop at the
Atlantic Mathematical Science Institute whose hospitality the author
acknowledges.

\appendix

\section{Quantum toroidal algebra}
\label{sec:prop-repr-quant}
In this Appendix we collect some properties of the quantum toroidal
algebra $U_{q,t}(\widehat{\widehat{\mathfrak{gl}}}_1)$.

\subsection{Generators}
\label{sec:generators-relations}

Quantum toroidal algebra
$U_{\mathfrak{q},\mathfrak{t}}(\widehat{\widehat{\mathfrak{gl}}}_1)$
is generated by PBW-type generators $P_{(n,m)}$, $(n,m) \in
\mathbb{Z}^2$ and a pair of central elements $C$, $C_{\perp}$. The
commutation relations are quite intricate and we will not write them
out here, instead referring to~\cite{Z5}. Instead we list some
important properties:
\begin{enumerate}
\item For trivial central charges $C = C_{\perp} = 1$ the commutation
  relations are invariant under $SL(2,\mathbb{Z})$ acting on the
  indices of the generators.

\item The generators corresponding to parallel integer vectors
  commute.

\item The algebra is doubly graded with gradings $d$, $d_{\perp}$
  acting as follows:
  \begin{equation}
    \label{eq:45}
    [d, P_{(n,m)}] = n P_{(n,m)}, \qquad      [d_{\perp}, P_{(n,m)}] = m P_{(n,m)}.
  \end{equation}

\item The algebra is invariant under any permutation of
  $\mathfrak{q}$, $\mathfrak{t}^{-1}$ and $\mathfrak{t}/\mathfrak{q}$
  parameters (but its representations usually are not).
\end{enumerate}


\subsection{Coproducts}
\label{sec:coproducts}

The algebra
$U_{\mathfrak{q},\mathfrak{t}}(\widehat{\widehat{\mathfrak{gl}}}_1)$
has an infinite number of coproducts
$\Delta_{\vartheta}:U_{\mathfrak{q},\mathfrak{t}}(\widehat{\widehat{\mathfrak{gl}}}_1)
\to U_{\mathfrak{q},\mathfrak{t}}(\widehat{\widehat{\mathfrak{gl}}}_1)
\hat{\otimes}
U_{\mathfrak{q},\mathfrak{t}}(\widehat{\widehat{\mathfrak{gl}}}_1)$ on
which $SL(2,\mathbb{Z})$ automorphism group acts transitively. The
coproduct is fixed by the choice of a Borel subalgebra
$\mathcal{B}_{\vartheta}$, which in turn is parametrized by a phase
$\vartheta \in [0,2\pi)$ such that $\mathrm{Arctg} (\vartheta)$ is
irrational. The phase defines a line in $\mathbb{R}^2$ separating the
$\mathbb{Z}^2$ plane of generators into two halves with
$\mathcal{B}_{\zeta}$ generated by $P_{(n,m)}$ with $(n,m)$ in the
left half (when looking in the direction of $\vartheta$).

Let us describe one of the coproducts explicitly. To this end we
introduce the generating currents
\begin{align}
  \label{eq:34}
  x^{\pm}(z) &= \sum_{n \in \mathbb{Z}} P_{(\pm 1,n)} z^{-n},\\
  \psi^{\pm}(z) &= C_{\perp}^{\pm \frac{1}{2}}\exp \left[ -\sum_{n \geq 1} \frac{\kappa_n}{n}
    P_{(0, \pm n)} z^{\mp n} \right],
\end{align}
where
\begin{equation}
  \label{eq:35}
  \kappa_n = (1-\mathfrak{q}^n)(1-\mathfrak{t}^{-n})\left( 1 - \left( \frac{\mathfrak{t}}{\mathfrak{q}} \right)^n \right).
\end{equation}
The coproduct $\Delta_{-\frac{\pi}{2}-\epsilon}$ corresponding to
$\mathcal{B}_{-\frac{\pi}{2}-\epsilon} = \left\langle e_{(n,m)}| n>0
  \text{ or } n = 0, m>0 \right\rangle$ on the generating currents is
given by
\begin{align}
\label{eq:36}  \Delta_{-\frac{\pi}{2}-\epsilon}(x^{+}(z)) &= x^{+}(z) \otimes 1 +
  \psi^{-}\left(C_{(1)}^{\frac{1}{2}}
    z\right) \otimes x^{+}(C_{(1)} z),\\
  \Delta_{-\frac{\pi}{2}-\epsilon}(x^{-}(z)) &=x^{-}\left(C_{(2)}
    z\right) \otimes \psi^{+}\left( C_{(2)}^{\frac{1}{2}}z
  \right) + 1 \otimes
  x^{-}(z),\label{eq:37}\\
  \Delta_{-\frac{\pi}{2}-\epsilon}(\psi^{\pm}(z)) &= \psi^{\pm}\left(
    C_{(2)}^{\pm \frac{1}{2}}z \right) \otimes \psi^{\pm}\left(
    C_{(1)}^{\mp \frac{1}{2}}z \right),\label{eq:38}\\
  \Delta_{-\frac{\pi}{2}-\epsilon}(C) &= C \otimes
  C \label{eq:39}.
\end{align}

\subsection{Drinfeld twists and universal $R$-matrices}
\label{sec:-drinfeld-twists}
As we have mentioned in sec.~\ref{sec:coproducts} different coproducts
are related by $SL(2,\mathbb{Z})$ automorphisms of the algebra. More
explicitly we have
\begin{align}
  \label{eq:31}
  (\mathcal{T}\otimes \mathcal{T})
  \Delta_{\vartheta}(\mathcal{T}^{-1}(g)) &= \Delta_{\mathrm{Arctg}(\tan
    \vartheta
    + 1)}(g),\\
  (\mathcal{S}\otimes \mathcal{S})
  \Delta_{\vartheta}(\mathcal{S}^{-1}(g)) &= \Delta_{\vartheta+
    \frac{\pi}{2}}(g)
\end{align}
for any $g \in
U_{\mathfrak{q},\mathfrak{t}}(\widehat{\widehat{\mathfrak{gl}}}_1)$,
where $\mathcal{S} = \left(
  \begin{smallmatrix}
    0&1\\
    -1&0
  \end{smallmatrix}
\right)$, $\mathcal{T}= \left(
  \begin{smallmatrix}
    1& 0\\
    1 &1
  \end{smallmatrix}
\right)$ are the generators of $SL(2,\mathbb{Z})$.  In other words,
$SL(2,\mathbb{Z})$ acts naturally on the set of directed lines with
irrational slopes in $\mathbb{R}^2$ plane, hence on the set of
coproducts. In particular, the ``vertical'' coproducts
$\Delta_{-\frac{\pi}{2}\pm \epsilon}$ are $\mathcal{T}$-invariant:
\begin{equation}
  \label{eq:32}(\mathcal{T}\otimes \mathcal{T})
    \Delta_{-\frac{\pi}{2}\pm \epsilon}(\mathcal{T}^{-1}(g)) =
    \Delta_{-\frac{\pi}{2}\pm \epsilon}(g)
\end{equation}
More generally, the coproduct $\Delta_{\mathrm{Arctg} \left( \frac{b}{a}
  \right) \pm \epsilon}$ is invariant under the subgroup of
$SL(2,\mathbb{Z})$ generated by
\begin{equation}
  \label{eq:33}
  \left(
    \begin{array}{cc}
      1+ab & -a^2\\
      b^2 & 1-ab
    \end{array}
\right).
\end{equation}

Crucially for our amalysis of wall-crossing, coproducts for different
slopes $\vartheta$ are also related to each other by nontrivial
Drinfeld twists $F_{\vartheta, \vartheta'} \in
U_{\mathfrak{q},\mathfrak{t}}(\widehat{\widehat{\mathfrak{gl}}}_1)
\hat{\otimes} U_{\mathfrak{q},\mathfrak{t}}(\widehat{\widehat{\mathfrak{gl}}}_1)$. We
have
\begin{equation}
  \label{eq:40}
  \Delta_{\vartheta'}(g) = F^{-1}_{\vartheta,\vartheta'} \Delta_{\vartheta}(g)F_{\vartheta,\vartheta'}
\end{equation}
for an element $g \in
U_{\mathfrak{q},\mathfrak{t}}(\widehat{\widehat{\mathfrak{gl}}}_1)$. There
is an explicit expression for $F_{\vartheta, \vartheta'}$~\cite{FJMM}:
\begin{equation}
  \label{eq:41}
  F_{\vartheta,\vartheta'} = \prod^{\to}_{
      \begin{smallmatrix}
        \gcd(a,b)=1\\
        \vartheta < \mathrm{Arctg} (\frac{b}{a}) < \vartheta'
      \end{smallmatrix}
    }
    F_{\mathrm{Arctg} (\frac{b}{a})},
\end{equation}
where the product is taken over all rational slopes between
$\vartheta$ and $\vartheta'$ in the order of increasing
$\mathrm{Arctg} (\frac{a}{b})$ (understood as a multivalued function)
and an ``elementary twist'' $F_{\mathrm{Arctg} (\frac{b}{a})}$
corresponding to a rational slope $\frac{b}{a}$ is given by
\begin{equation}
  \label{eq:42}
  F_{\mathrm{Arctg} (\frac{b}{a})} = \exp \left[ \sum_{n \geq 1} \frac{\kappa_n}{n} P_{(na,nb)}
    \otimes P_{(-na,-nb)} \right].
\end{equation}

The Drinfeld twists thus defined are multiplicative in $\vartheta$:
\begin{equation}
  \label{eq:43}
  F_{\vartheta,\vartheta'} F_{\vartheta',\vartheta''} =
  F_{\vartheta,\vartheta''}  
\end{equation}

We denote the universal $R$-matrix for the coproduct $\Delta_{\vartheta}$
by $\mathcal{R}_{\vartheta} \in
U_{\mathfrak{q},\mathfrak{t}}(\widehat{\widehat{\mathfrak{gl}}}_1)
\hat{\otimes}
U_{\mathfrak{q},\mathfrak{t}}(\widehat{\widehat{\mathfrak{gl}}}_1)$,
so that
\begin{equation}
  \label{eq:46}
  \Delta_{\vartheta}^{\mathrm{op}}(g) = \mathcal{R}_{\vartheta}
  \Delta_{\vartheta}(g) \mathcal{R}_{\vartheta}^{-1}
\end{equation}
for any $g \in
U_{\mathfrak{q},\mathfrak{t}}(\widehat{\widehat{\mathfrak{gl}}}_1)$. The
universal $R$-matrix is essentially a twist corresponding to a $\pi$
rotation of the coproduct~\cite{N}:
\begin{equation}
  \label{eq:44}
  \mathcal{R}_{\vartheta} = P e^{c \otimes d + d \otimes c + c_{\perp}
    \otimes d_{\perp} + d_{\perp} \otimes c_{\perp}} F_{\vartheta, \vartheta+ \pi}.
\end{equation}
where $P$ is a permutation of tensor factors, $c = \ln C$, $c_{\perp}
= \ln C_{\perp}$ and $d$, $d_{\perp}$ are the two gradings. The
product expression~\eqref{eq:41} for the twist in the case of the
universal $R$-matrix is known as the Khoroshkin-Tolstoy
formula~\cite{KT1,KT2}.
  
\subsection{Vector representations}
\label{sec:vect-repr}
There is a representation $\rho_{\mathcal{V}_\mathfrak{q}^{*}}$ of
$U_{\mathfrak{q},\mathfrak{t}}(\widehat{\widehat{\mathfrak{gl}}}_1)$
using a pair variables $\mathbf{x}$ and $\mathbf{y}$ satisfying
$\mathfrak{q}$-commutation relations~\eqref{eq:8}. We have
\begin{align}
  \rho_{\mathcal{V}^{*}_{\mathfrak{q}}} (C) &= \rho_{\mathcal{V}^{*}_{\mathfrak{q}}} (C_{\perp}) =  1,\\
  \rho_{\mathcal{V}^{*}_{\mathfrak{q}}}(P_{(n,m)}) &=
  \frac{\mathfrak{q}^{-\frac{nm}{2}}}{\mathfrak{q}^{\mathrm{gcd(n,m)}/2}-\mathfrak{q}^{-\mathrm{gcd(n,m)}/2}}
  \mathbf{x}^m \mathbf{y}^{-n},   \label{eq:47}
\end{align}
where $\mathrm{gcd}(n,m)$ denotes the greatest common divisor of $n$
and $m$ which we understand to be always positive. From
Eq.~\eqref{eq:47} we get the expressions for the generating currents:
\begin{align}
  \rho_{\mathcal{V}^{*}_{\mathfrak{q}}}(x^{\pm}(z)) &=
  \frac{1}{\mathfrak{q}^{1/2} - \mathfrak{q}^{-1/2}} \delta \left(
    \mathfrak{q}^{\mp 1/2} \frac{\mathbf{x}}{z} \right)
  \mathbf{y}^{\mp 1},\\
  \rho_{\mathcal{V}^{*}_{\mathfrak{q}}}(\psi^{+}(z)) &= \frac{\left( 1
      - \frac{t}{\sqrt{\mathfrak{q}}} \frac{\mathbf{x}}{z} \right)
    \left( 1 - \frac{\sqrt{\mathfrak{q}}}{\mathfrak{t}}
      \frac{\mathbf{x}}{z} \right)}{\left( 1 - \sqrt{\mathfrak{q}}
      \frac{ \mathbf{x}}{z} \right) \left( 1 -
      \frac{1}{\sqrt{\mathfrak{q}}} \frac{\mathbf{x}}{z} \right)},\\
  \rho_{\mathcal{V}^{*}_{\mathfrak{q}}}(\psi^{-}(z)) &= \frac{\left( 1
      - \frac{\sqrt{\mathfrak{q}}}{\mathfrak{t}} \frac{z}{\mathbf{x}}
    \right) \left( 1 - \frac{\mathfrak{t}}{\sqrt{\mathfrak{q}}}
      \frac{z}{ \mathbf{x}} \right)}{\left( 1 -
      \frac{1}{\sqrt{\mathfrak{q}}} \frac{z}{ \mathbf{x}}
    \right) \left( 1 - \sqrt{\mathfrak{q}} \frac{z}{\mathbf{x}} \right)}.
\end{align}
Let us note that $\rho_{\mathcal{V}^{*}_{\mathfrak{q}}}(\psi^{\pm}(z))$ are
expansions of the same rational function in positive or negative
powers of $z$ respectively.

\paragraph{Tensor product of vector representations.}
\label{sec:tens-prod-vect}
When tensoring representations of
$U_{\mathfrak{q},\mathfrak{t}}(\widehat{\widehat{\mathfrak{gl}}}_1)$
one always needs to specify which coproduct is used. We will use the
notation $\mathcal{V}_{\mathfrak{q}}^{*} \otimes_{\vartheta} \mathcal{V}_{\mathfrak{q}}^{*}$ for a tensor product
with the action of the algebra defined by $\Delta_{\vartheta}$. Here
we give some explicit formulas for the action of the generating
currents on the tensor product:
\begin{multline}
  \label{eq:48}
  \rho_{\mathcal{V}_{\mathfrak{q}}^{*}
    \otimes_{-\frac{\pi}{2}-\epsilon} \mathcal{V}_{\mathfrak{q}}^{*}}
  (P_{(1,n)}) = \rho_{\mathcal{V}_{\mathfrak{q}}^{*}} \otimes
  \rho_{\mathcal{V}_{\mathfrak{q}}^{*}}
  (\Delta_{-\frac{\pi}{2}-\epsilon}(x^{+}(z))) =\\
  = \frac{\mathfrak{q}^{- n/2} }{\mathfrak{q}^{1/2} - \mathfrak{q}^{-1/2}} \left(   \mathbf{x}_1^n
    \mathbf{y}_1^{-1} + \frac{\left( 1 - \frac{1}{\mathfrak{t}}
        \frac{\mathbf{x}_2}{\mathbf{x}_1} \right) \left( 1 -
        \frac{\mathfrak{t}}{\mathfrak{q}} \frac{\mathbf{x}_2}{
          \mathbf{x}_1} \right)}{\left( 1 - \frac{1}{\mathfrak{q}}
        \frac{\mathbf{x}_2}{ \mathbf{x}_1} \right) \left( 1 -
        \frac{\mathbf{x}_2}{\mathbf{x}_1} \right)} 
      \mathbf{x}_2^n
    \mathbf{y}_2^{-1}\right)
\end{multline}
\begin{multline}
  \label{eq:5}
  \rho_{\mathcal{V}_{\mathfrak{q}}^{*}
    \otimes_{-\frac{\pi}{2}-\epsilon} \mathcal{V}_{\mathfrak{q}}^{*}}
  (P_{(-1,n)}) = \rho_{\mathcal{V}_{\mathfrak{q}}^{*}} \otimes
  \rho_{\mathcal{V}_{\mathfrak{q}}^{*}}
  (\Delta_{-\frac{\pi}{2}-\epsilon}(x^{-}(z))) =\\
  = \frac{\mathfrak{q}^{n/2}}{\mathfrak{q}^{1/2} -
    \mathfrak{q}^{-1/2}} \left( \frac{\left( 1 -
        \frac{1}{\mathfrak{t}} \frac{\mathbf{x}_1}{\mathbf{x}_2}
      \right) \left( 1 - \frac{\mathfrak{t}}{\mathfrak{q}}
        \frac{\mathbf{x}_1}{ \mathbf{x}_2} \right)}{\left( 1 -
        \frac{1}{\mathfrak{q}} \frac{\mathbf{x}_1}{ \mathbf{x}_2}
      \right) \left( 1 - \frac{\mathbf{x}_1}{\mathbf{x}_2} \right)}
    \mathbf{x}_1^n \mathbf{y}_1 + \mathbf{x}_2^n \mathbf{y}_2\right)
\end{multline}
\begin{equation}
  \label{eq:14}
    \rho_{\mathcal{V}_{\mathfrak{q}}^{*}
    \otimes_{-\frac{\pi}{2}-\epsilon} \mathcal{V}_{\mathfrak{q}}^{*}}
  (P_{(0,m)}) = \rho_{\mathcal{V}_{\mathfrak{q}}^{*}} \otimes
  \rho_{\mathcal{V}_{\mathfrak{q}}^{*}}
  (\Delta_{-\frac{\pi}{2}-\epsilon}(P_{(0,m)})) =
  \frac{1}{\mathfrak{q}^{m/2} - \mathfrak{q}^{-m/2}} \left(
    \mathbf{x}_1^m + \mathbf{x}_2^m \right)
\end{equation}
where $\mathbf{x}_i, \mathbf{y}_i$, $i=1,2$ are
$\mathfrak{q}$-commuting operators from the first or second vector
representation.

\paragraph{$SL(2,\mathbb{Z})$ action in $\mathcal{V}^{*}_{\mathfrak{q}}$.}
\label{sec:sl2-z-action}
The group $SL(2,\mathbb{Z})$ acts by inner automorphisms in the vector
representation $\mathcal{V}^{*}_{\mathfrak{q}}$. Physically this
follows from the fact that $S$-duality leaves the D3 brane
invariant. The action of $SL(2,\mathbb{Z})$ is as follows:
\begin{align}
  \label{eq:49}
  \rho_{\mathcal{V}^{*}_{\mathfrak{q}}}(\mathcal{T}(P_{(n,m)})) &=
  e^{\frac{(\ln \mathbf{x})^2}{2 \ln q}}
  \rho_{\mathcal{V}^{*}_{\mathfrak{q}}}(P_{(n,m)}) e^{-\frac{(\ln
      \mathbf{x})^2}{2 \ln q}},\\
    \rho_{\mathcal{V}^{*}_{\mathfrak{q}}}(\mathcal{S}(P_{(n,m)})) &=
  e^{-\frac{\ln \mathbf{x} \ln \mathbf{y}}{ \ln q}}
  \rho_{\mathcal{V}^{*}_{\mathfrak{q}}}(P_{(n,m)}) e^{\frac{\ln \mathbf{x} \ln \mathbf{y}}{ \ln q}}.
\end{align}

\end{document}